\documentstyle[prl,aps,epsf,floats]{revtex}
\begin{document}
\draft
\twocolumn[\hsize\textwidth\columnwidth\hsize\csname @twocolumnfalse\endcsname
\title{Evolution of Hole and Spin Dynamics 
       in High T$_{c}$ Superconductors \\
       within the Small Hole Density Limit 
       of the $t-J$ Model}    
\author{Bumsoo Kyung}
\address{Max Planck Institute for Physics of Complex Systems, 
         Noethnitzer Str. 38, 01187 Dresden, Germany}
\date{2 October 1997}
\maketitle
\begin{abstract}

   The evolution of hole and spin dynamics 
in high T$_{c}$ superconductors is studied  
within the self-consistent noncrossing approximation of the $t-J$ model
in the small hole density limit. 
As the doping concentration is increased,
long-range electron correlations disappear rapidly and 
the quasiparticle
energy band becomes considerably narrow.
At a small hole density
long-range antiferromagnetic order is destroyed leading to the 
inadequacy of spin wave basis approximation near small 
wave vectors. Spin excitations
near the antiferromagnetic zone boundary are strongly  
renormalized and damped but they are still well described within
spin wave basis approximation.
\end{abstract}
\vskip2pc]
\narrowtext
%
%
%
   
   Since the discovery of high temperature superconductivity in the 
copper oxide compounds, spin dynamics in those materials has received
considerable attention in an attempt to understand various anomalous
properties of the normal state as well as superconductivity itself.
In the cuprates spin fluctuations are believed as a dominant  
interaction channel 
between charge carriers,
because strong electron correlations present in the copper oxide planes
tend to significantly suppress charge fluctuations. 
In this respect, recently the strong correlation based 
Hubbard and $t-J$ models have been intensively
studied by various means%
\cite{Dagotto:1994} in order to study spin dynamics in cuprates.
At half-filling,   
the quantum Heisenberg model%
\cite{Chakravarty:1988}
has exhibited both 
qualitative
and quantitative agreement with most of the experimental findings on undoped 
(parent) compounds.
Spin dynamics away from half-filling, however, is still far from 
understanding, although many anomalous properties and even 
superconductivity have been observed in this limit.
This lack of satisfactory microscopic models away from half-filling
has lead to introducing various phenomenological models%
\cite{Millis:1990,Kampf:1990}.
In the present paper we extend the single hole problem
studied previously by various authors%
\cite{Schmitt-Rink:1988},
to the case with a finite but small density of doped holes
within the well-established
self-consistent noncrossing approximation.
As a result of the self-consistent nature of the calculation,
hole dynamics is also studied on the equal footing.
Our aim here is to examine the evolution of hole and spin dynamics
in the small doping regime from the insulating (undoped) side.

   The $t-J$ Hamiltonian reads
\begin{equation}
  H_{t-J} = -t\sum_{<i,j>,\sigma}\hat{c}^{+}_{i,\sigma}
               \hat{c}_{j,\sigma}
               + J\sum_{<i,j>}\vec{S}_{i}\cdot
               \vec{S}_{j} \; ,
                                                          \label{eq1}
\end{equation}
where $J$ is the spin exchange constant.
The $\hat{c}^{+}_{i,\sigma}$ and $\hat{c}_{i,\sigma}$ are electron
creation and annihilation operators acting in the reduced Hilbert
space without double occupancy at the same site. 
After Fourier and Bogoliubov transformations to Eq.~(\ref{eq1}),
the $t-J$ model Hamiltonian becomes  
within the spin wave basis approximation%
\cite{Schmitt-Rink:1988},
\begin{eqnarray}
H_{t-J} \; = \; \frac{1}{\sqrt{N}}\sum_{\vec{k},\vec{q}}
           h^{+}_{\vec{k}}h_{\vec{k}
         - \vec{q}}(M_{\vec{k},\vec{q}}\alpha_{\vec{q}}  
         + N_{\vec{k},\vec{q}}\alpha^{+}_{-\vec{q}})  
                                                            \nonumber   \\
         + \sum_{\vec{q}}\Omega_{\vec{q}}\alpha^{+}_{\vec{q}}  
           \alpha_{\vec{q}}  
         - \mu\sum_{\vec{k}}h^{+}_{\vec{k}}h_{\vec{k}} \; ,  \label{eq2}
\end{eqnarray}
where $h_{\vec{k}}$ and
$\alpha_{\vec{q}}$ denote 
a hole annihilation operator
with momentum $\vec{k}$ and 
a spin wave annihilation operator with momentum $\vec{q}$, respectively.
$M_{\vec{k},\vec{q}}$ and $N_{\vec{k},\vec{q}}$ are 
the bare interaction vertices, and 
$\Omega_{\vec{q}}$ is the spin wave energy with momentum $\vec{q}$.
See Refs.~\onlinecite{Schmitt-Rink:1988} for details.
$\mu$ is the chemical
potential controlling the doping concentration.
The hopping term of the $t-J$ model transforms into a
coupling term proportional to $t$ due to the on-site no
double occupancy constraint.
Since spins are ordered and thus spin waves are well defined excitations
within the spin-spin correlation length,  
linear spin wave approximation is expected to be good for a small 
density limit of doped holes which the present paper is focused on.

   In this paper we employ the self-consistent noncrossing 
approximation in the electron self-energy and spin excitation 
polarization functions.
The vertex correction neglected in this approximation was shown 
by Sherman\cite{Sherman:1993} to be of order of the doping 
concentration so that it is quite a reliable approximation for the 
present study.
Within the self-consistent noncrossing approximation,
the hole self-energy and 
the $2 \times 2$ matrix
spin polarization functions are given in terms of 
a $2 \times 2$ matrix spin excitation propagator $D(q)$ and 
a two component interaction vertex $M$,
\begin{eqnarray}
 \Sigma(k) & = & - \frac{T}{N}\sum_{q} 
                M^{+}D(q)M G(k-q)     \; ,
                                                      \nonumber  \\
 \Pi(q) & = & \frac{T}{N}\sum_{k} 
          M M^{+} G(k)G(k-q)
                                               \; ,
                                                           \label{eq6}
\end{eqnarray}
where $T$ is the temperature and 
$M$ is defined by  
\begin{equation}
  M = \left( \begin{array}{c}
                   M_{\vec{k},\vec{q}} \\
                   N_{\vec{k},\vec{q}} 
              \end{array}
      \right) \; .
                                                         \label{eq7}
\end{equation}
Here $k$ and $q$ are compact notations for $(\vec{k},ik_{n})$ and 
$(\vec{q},iq_{n})$, respectively, where $k_{n}$ and $q_{n}$
are Fermionic and Bosonic Matsubara frequencies.
Through the hole self-energy and spin polarizations calculated
above, we obtain the interacting hole and spin excitation Green's functions
after taking the analytic continuation to get retarded functions,
\begin{eqnarray}
G^{R}(\vec{k},\omega) \; = \; \frac{1}{\omega-\Sigma^{R}(\vec{k},
                            \omega)+\mu+i\delta} \; ,     
                                                        \nonumber     \\
D^{R}(\vec{q},\nu) \; = \; \frac{1}{[D^{0,R}(\vec{q},\nu)]^{-1}-\Pi^{R}
                         (\vec{q},\nu)} \; ,                \label{eq8}
\end{eqnarray}
where the inverse noninteracting spin wave propagator 
$[D^{0,R}(\vec{q},y)]^{-1}$ is given as
\begin{equation}
[D^{0,R}(\vec{q},\nu)]^{-1} \; = \; 
             \left( \begin{array}{cc}
                    \nu-\Omega_{\vec{q}}+i\delta & 0
                                          \\
                    0 & -\nu-\Omega_{\vec{q}}-i\delta 
                    \end{array} 
             \right) \; .
                                                         \label{eq9}
\end{equation}
Because of its matrix structure,
$D(q)$ contains four terms associated with two diagonal (normal) and 
two off-diagonal (anomalous) components.
Here we take spin wave excitations as the zeroth order or unperturbed
states as seen in Eq.~(\ref{eq8})
and the present calculation can deal with the renormalization and damping 
of the spin excitation modes by taking into account spin polarizations, 
$\Pi_{ij}^{R}(\vec{q},\nu)$, in the spin excitation Green's function.
Furthermore, a possible inadequacy of the spin wave basis approximation
in some region of the Brillouin zone
may be gauged by the appearance of the off-diagonal components of the  
matrix spin excitation propagator, which will be discussed later.
It should be noted that in Eqs.~(\ref{eq8}) and (\ref{eq9}) 
we did not attach any artificial damping
factor in the denominator of both the hole and spin excitation 
Green's functions in order to obtain the correct hole spectral functions
without smearing of any fine structures.
Eqs.~(\ref{eq6}) and (\ref{eq8}) are iterated
until required convergency is reached.
In this way both the hole and spin excitation Green's functions are 
computed self-consistently.
Throughout the paper,
$J/t=0.4$ and the zero temperature limit ($T=0$) are used and
a lattice of $20 \times 20$ is used to discretize momentum space.
Recently Sherman\cite{Sherman:1997}
used similar equations to study the doping dependence of hole and 
spin dynamics in cuprates.

\begin{figure}
 \vbox to 7.0cm {\vss\hbox to -5.0cm
 {\hss\
       {\includegraphics{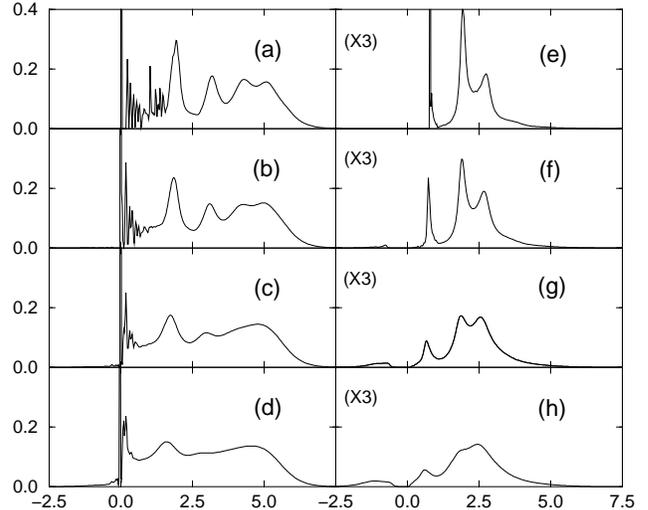}
       }
  \hss}
 }
\caption{Hole spectral functions at $(\pi/2,\pi/2)$  
         and at $(0,0)$ for four different 
         doping concentrations.
         (a), (b), (c) and (d) are the hole spectral functions 
         at $(\pi/2,\pi/2)$
         for $0.0$, $0.95$, $2.93$ and $5.54 \%$,
         respectively.
         (e), (f), (g) and (h) are the hole spectral functions at $(0,0)$
         for $0.0$, $0.95$, $2.93$ and $5.54 \%$,
         respectively, where the vertical axis is multiplied by 
         three to that for (a)-(d).
             }
\label{fig1}
\end{figure}

   Figures~\ref{fig1} show the evolution of the hole spectral functions
at $(\pi/2,\pi/2)$ and at $(0,0)$
for doping concentrations $0.0$, $0.95$, $2.93$
and $5.54 \%$. 
Due to the finite size of the lattice ($20 \times 20$)
in the Brillouin
zone we could not vary the hole density in a continuous manner but 
instead in a discrete way as indicated above.
The most striking change of the spectral functions at
$(\pi/2,\pi/2)$ by doping is the significant suppression of string 
excitations. For $5.54 \%$ hole doping  
(Fig.~\ref{fig1}(d)) 
these multiple magnon excitations
become strongly damped and are absorbed into a featureless continuous
background.
Since higher string excitations correspond to longer-range correlations,
they are more susceptible to doping, as seen in the figure.
The strong suppression of the string excitations 
for $5.54 \%$ doping concentration
implies that the effective hopping
rate $t_{eff}$ 
already becomes comparable to $J$, leading to 
the considerable loss of 
strong correlations even at this low doping.

\begin{figure}
 \vbox to 7.0cm {\vss\hbox to -5.0cm
 {\hss\
       {\includegraphics{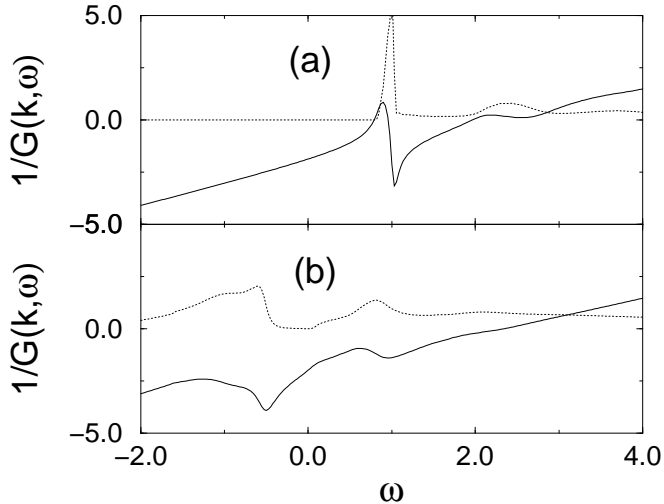}
       }
  \hss}
 }
\caption{Inverse of the interacting Green's functions at $(0,0)$ for two 
         different doping concentrations.
         The solid and dotted curves denote the real and imaginary
         parts, respectively.
         (a) The doping concentration $0.0 \%$.
         (b) The doping concentration $5.54 \%$.
             }
\label{fig2}
\end{figure}

   The hole spectral functions at $(0,0)$ are shown 
in Figs.~\ref{fig1}(e)-(h)
for the same doping concentrations as in Figs.~\ref{fig1}(a)-(d).
Although the spectral weight of the lowest peak at $(0,0)$ is 
relatively smaller than that at $(\pi/2,\pi/2)$ for $x=0.0 \%$ 
a well-defined  
quasiparticle state with an infinite lifetime exists near the bottom of 
the spectrum
(Fig.~\ref{fig1}(e))
at $(0,0)$ point.
For only a few percent of doping concentration, however, 
the quasiparticle state at this momentum
is completely destroyed and the incoherent part becomes more dominating.
This can be easily understood by examining the real and
imaginary parts of the inverse hole Green's
function as exhibited in Figs.~\ref{fig2}.
While the pole condition (vanishing real and imaginary parts of the inverse
Green's function at the same frequency)
is satisfied nearly at $2J$ or $0.8t$ for the undoped case
(Fig.~\ref{fig2}(a)),
just a few percent doping changes both the real and  
imaginary parts so drastically near $2J$ that the quasiparticle  
condition is never 
met as shown in Fig.~\ref{fig2}(b).
On the other hand, the change of the self-energy at $(\pi/2,\pi/2)$ 
(which is not shown here)
is much smaller than that at the origin 
and thus the quasiparticle state persists at $(\pi/2,\pi/2)$. 
The disappearance of the quasiparticle state by doping is found 
in the significant portion of the Brillouin zone near the origin.
This is mainly responsible for the dynamical narrowing of the quasiparticle
band by doping,
since the quasiparticle states near the origin 
comprise the large portion of 
the upper part of the quasiparticle band at $x=0.0$.
The total quasiparticle bandwidth is approximately $2J$ for $0.0 \%$
doping, while it is reduced to less than $0.8J$ for $5.54 \%$ doping
by dynamical effects.
We also found that doping not only makes high lying states near $(0,0)$
point unstable, but also strongly modifies low lying states.
As the doping concentration is increased, the Fermi surfaces centered 
at $(\pm\pi/2,\pm\pi/2)$
with an ellipsoidal shape
become so rapidly elongated along the antiferromagnetic zone boundary 
that they become a large (connected) Fermi surface about $10 \%$
doping level
(Details will be published elsewhere\cite{Kyung:1997-2}).

\begin{figure}
 \vbox to 7.0cm {\vss\hbox to -5.0cm
 {\hss\
       {\includegraphics{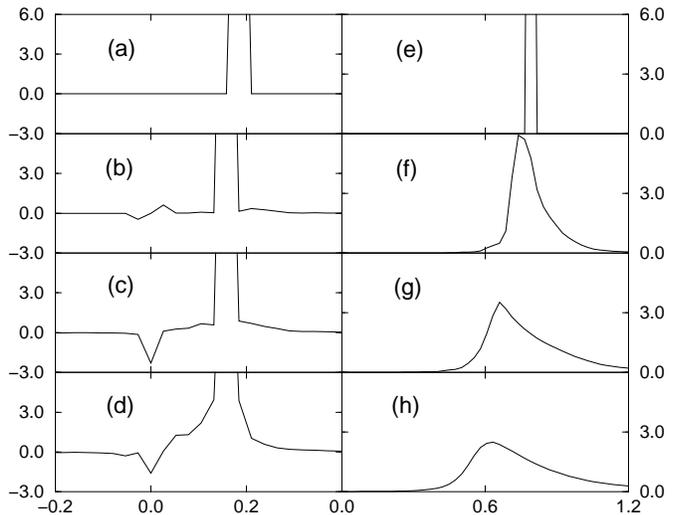}
       }
  \hss}
 }
\caption{Spectral functions of the diagonal component of the matrix 
         spin excitation propagator, $B_{11}(\vec{q},\nu)$,
         at $(0,\pi/10)$ and at $(\pi/2,\pi/2)$ for four different 
         doping concentrations.
         (a), (b), (c) and (d) are $B_{11}(\vec{q},\nu)$ at $(0,\pi/10)$
         for $0.0$, $0.95$, $2.93$ and $5.54 \%$,
         respectively.
         (e), (f), (g) and (h) are $B_{11}(\vec{q},\nu)$ at $(\pi/2,\pi/2)$
         for $0.0$, $0.95$, $2.93$ and $5.54 \%$,
         respectively.
             }
\label{fig3}
\end{figure}

\begin{figure}
 \vbox to 7.0cm {\vss\hbox to -5.0cm
 {\hss\
       {\includegraphics{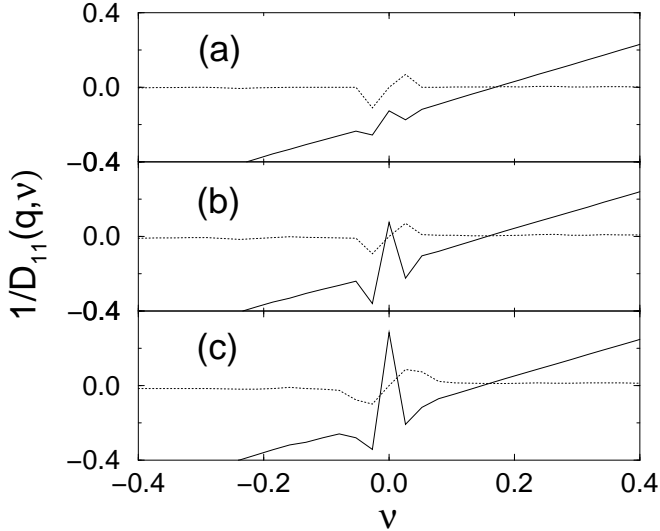}
       }
  \hss}
 }
\caption{Inverse of the diagonal component of the matrix 
         spin excitation propagator
         at $(0,\pi/10)$ for three different 
         doping concentrations.
         The solid and dotted curves denote the real and imaginary
         parts, respectively.
         (a) The doping concentration $0.95 \%$.
         (b) The doping concentration $2.93 \%$.
         (c) The doping concentration $5.54 \%$.
             }
\label{fig4}
\end{figure}

   Figures~\ref{fig3}(a)-(d) show $B_{11}(\vec{q},\nu)$ at the smallest
momentum $(0,\pi/10)$ in the present calculation
for several doping concentrations.
In the undoped case spin wave excitations are eigenstates of our
model Hamiltonian where they have an infinitely long lifetime as seen in 
Fig.~\ref{fig3}(a). 
For 0.95, 2.93, 5.54 $\%$ doping concentrations, the renormalization of
the peak strength is small and we find only their small downward shifts,
namely, softening of the spin excitation velocity.
For 2.93 $\%$ doping (Fig.~\ref{fig3}(c)), 
however, a pronounced peak occurs near the 
origin, although its spectral weight is relatively smaller than
that of the main peak.
The occurrence of this peak can be understood more clearly by examining the 
real and imaginary parts of the inverse $D_{11}(\vec{q},\nu)$, as 
shown in Figs.~\ref{fig4}.
For 0.95 $\%$ concentration (Fig.~\ref{fig4}(a)),
the pole condition is fulfilled only at
$0.16t$ and some small structure due to doping develops near the origin. 
For 2.93 $\%$ doping
(Fig.~\ref{fig4}(b)),
this new structure grows so drastically that the pole condition is also
satisfied near the origin.
Its small spectral weight is caused by its small residue
at that frequency.
At 5.54 $\%$ doping 
(Fig.~\ref{fig4}(c)),
the real part grows further and 
the pole condition is less fully satisfied near the origin.
Thus, the new structure near the origin becomes smaller for 
higher than 2.93 $\%$ doping concentrations.
Evidently the appearance of the new mode near the origin happens at
between $0.95\%$ and $2.93\%$ doping within our numerical resolution.
This corresponds to the disappearance of antiferromagnetic 
long-range order, as first pointed out by 
Sherman {\em et al.}\cite{Sherman:1993}
These authors argued that 
the appearance of zero or negative frequency boson modes 
at zero temperature yields an
{\it infinite} number of bosonic excitations,  
implying the inadequacy of the spin wave basis 
approximation in that momentum.
It indicates the disappearance of long-range antiferromagnetic 
order.
This can be also gauged by the behavior of the anomalous (off-diagonal)
spin excitation propagators. 
We found that the off-diagonal spin propagators become
significant near the origin, $(0,0)$, while they are vanishingly 
small near the zone boundary
in spite of their strong renormalization and damping.
This means the condensation of long wavelength
spin excitation
modes, just as electrons condensate in the superconducting  
state. Apparently this is an unphysical result, caused by the spin wave 
basis approximation for that particular momentum.

     The main spin excitation peak 
in Fig.~\ref{fig3}(b) 
is slightly shifted downward for
0.95 $\%$ doping,
leading to the softening of the mode.  
We evaluated the concentration where the peak occurs at zero frequency 
by employing a simple interpolation scheme to higher doping cases, which 
is essentially the same method as 
Igarashi {\it et al.}\cite{Igarashi:1992} used. 
These authors calculated the amount of softening (shift) of the mode
by explicitly taking into account coherent and incoherent contributions
and also by using perturbation theory to obtain the spectral function 
for a finite density of holes
from that for a single hole. 
According to our simple interpolation, the complete softening of the main 
peak happens approximately at $11 \%$
in good agreement with $12\%$ found by
the above authors.
This is evidently too high to be reconciled with experiments.
The present calculation shows that the disappearance of long-range  
antiferromagnetic order is not determined by the complete softening of the 
mode but instead by the condition that the  
velocity of 
the long wavelength spin excitation 
becomes smaller than the
Fermi velocity.
As doping concentration is increased, the spin excitation velocity decreases
and at the same time the Fermi velocity rapidly increases 
as $\sqrt{x}$
in the small doping limit. 
Hence, self-consistent treatment of hole and spin dynamics is very 
important to predict a small value of the critical concentration.
The softening and damping of the spin excitation peaks at 
high momenta near the antiferromagnetic zone 
boundary is quite drastic even for small doping, as exhibited in
Figs.~\ref{fig3}(e)-(h). In spite of this significant change in the
spectral properties of spin excitations at these momenta, 
the off-diagonal (anomalous)
components of the spin excitation propagator is vanishingly small
and thus spin wave basis approximation appears still valid, in 
agreement with the report by Sokol {\it et al.}\cite{Sokol:1993}

   In summary, the evolution of hole and spin dynamics  
in high T$_{c}$ superconductors 
was studied within the
self-consistent noncrossing approximation in the small 
density limit of doped holes.  
As the doping concentration 
is increased, string excitations rapidly disappear
implying significant loss of long-range electron correlations.
High energy states also lose their 
quasiparticle properties completely and thus the quasiparticle
energy band becomes considerably narrower than that of the undoped case.
The strong change of the low lying quasiparticle 
band caused by dynamical effect 
of doped holes by exchange of spin excitations  
shows a tendency that separated small hole pockets  
become a large Fermi surface at relatively small doping concentrations. 
The disappearance of long-range antiferromagnetic order is found
at between $0.95 \%$ and $2.93 \%$ doping levels
both in the diagonal and off-diagonal components of the spin  
excitation propagator.
The spin 
excitations at high momenta become strongly  
renormalized and damped by doping, yet they are still well described within
spin wave basis approximation in agreement with the previous report.

    We would like to thank Professors J. Igarashi and P. Fulde 
for useful discussions.
\end{document}